 \documentclass[sigconf]{acmart}

\AtBeginDocument{%
  }

\setcopyright{acmlicensed}

\copyrightyear{2025}
\acmYear{2025}
\setcopyright{acmlicensed}\acmConference[SIGIR '25]{Proceedings of the 48th International ACM SIGIR Conference on Research and Development in Information Retrieval}{July 13--18, 2025}{Padua, Italy}
\acmBooktitle{Proceedings of the 48th International ACM SIGIR Conference on Research and Development in Information Retrieval (SIGIR '25), July 13--18, 2025, Padua, Italy}
\acmDOI{10.1145/3726302.3730092}
\acmISBN{979-8-4007-1592-1/2025/07}
\usepackage{algorithm2e}
\usepackage{algorithmic}
\usepackage{amsmath} 
\usepackage{xcolor} 
\usepackage{color}
\usepackage[breakable]{tcolorbox}
\usepackage{booktabs}
\usepackage{booktabs}
\usepackage{multirow}
\usepackage[normalem]{ulem}
\useunder{\uline}{\ul}{}
\usepackage{enumitem}
\begin{document}

\title{\textbf{\textit{The Truth Becomes Clearer Through Debate!}} Multi-Agent \\ Systems with Large Language Models Unmask Fake News}

\author{Yuhan Liu}
\affiliation{
  \institution{Gaoling School of Artificial Intelligence,\\ Renmin University of China}
  \city{Beijing}
  \country{China}
}
\email{yuhan.liu@ruc.edu.cn}

\author{Yuxuan Liu}
\affiliation{
  \institution{Gaoling School of Artificial Intelligence,\\ Renmin University of China}
  \city{Beijing}
  \country{China}
}

\author{Xiaoqing Zhang}
\affiliation{
  \institution{Gaoling School of Artificial Intelligence,\\ Renmin University of China}
  \city{Beijing}
  \country{China}
}

\author{Xiuying Chen}
\authornote{Corresponding authors.}
\affiliation{
  \institution{Mohamed bin Zayed University of Artificial Intelligence}
  \city{Abu Dhabi}
  \country{UAE}
}

\author{Rui Yan}
\authornotemark[1]
\affiliation{%
  \institution{ Gaoling School of Artificial Intelligence, \\ Renmin University of China}
  \city{Beijing}
  \country{China}
}
\affiliation{%
  \institution{Wuhan University}
  \city{Wuhan}
  \country{China}}


\renewcommand{\shortauthors}{Yuhan Liu, Yuxuan Liu, Xiaoqing Zhang, Xiuying Chen, Rui Yan}

\begin{abstract}
 In today's digital environment, the rapid propagation of fake news via social networks poses significant social challenges. 
Most existing detection methods either employ traditional classification models, which suffer from low interpretability and limited generalization capabilities, or craft specific prompts for large language models (LLMs) to produce explanations and results directly,  failing to leverage LLMs' reasoning abilities fully.
Inspired by the saying that ``truth becomes clearer through debate,'' our study introduces a novel multi-agent system with LLMs named TruEDebate (TED) to enhance the \textit{interpretability} and \textit{effectiveness} of fake news detection. 
TED employs a rigorous debate process inspired by formal debate settings.
Central to our approach are two innovative components: the DebateFlow Agents and the InsightFlow Agents.
The DebateFlow Agents organize agents into two teams, where one supports and the other challenges the truth of the news. These agents engage in opening statements, cross-examination, rebuttal, and closing statements, simulating a rigorous debate process akin to human discourse analysis, allowing for a thorough evaluation of news content.
Concurrently, the InsightFlow Agents consist of two specialized sub-agents: the Synthesis Agent and the Analysis Agent. 
The Synthesis Agent summarizes the debates and provides an overarching viewpoint, ensuring a coherent and comprehensive evaluation.
The Analysis Agent, which includes a role-aware encoder and a debate graph, integrates role embeddings and models the interactions between debate roles and arguments using an attention mechanism, providing the final judgment.
Our extensive experiments on two datasets, ARG-EN and ARG-CN, demonstrate that the TED framework surpasses traditional methods across various metrics and, more importantly, enhances interpretable fake news detection by illuminating logical reasoning and structured debate processes leading to accurate conclusions.
We release our code to support Information systems that use structured debate within \textit{responsible} information systems for improved decision-making\footnote{https://github.com/LiuYuHan31/TED\_fake-news-detction}.
\end{abstract}



\maketitle

\section{Introduction}
\begin{figure}[htbp]
    \centering
    \includegraphics[width=1\linewidth]{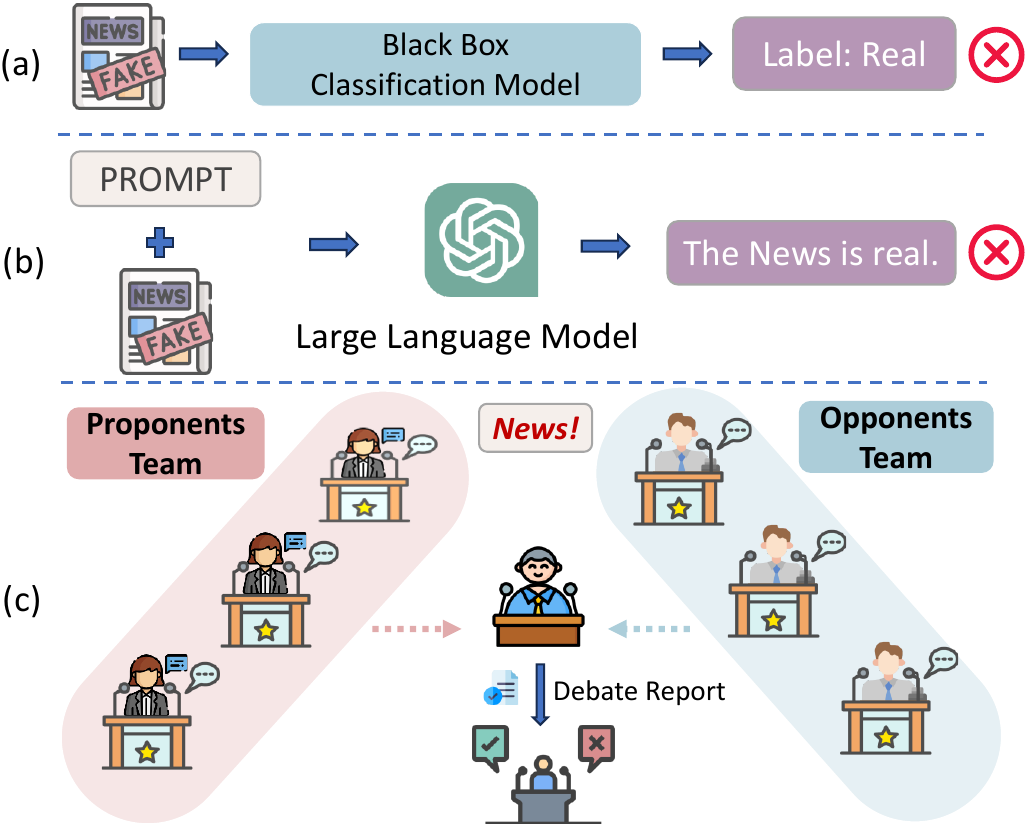}
    \caption{(a) Traditional models primarily rely on black box classification models for detecting fake news.
(b) LLMs are prompted to output detection results directly, which do not fully utilize the potential of large model reasoning.
(c) Our multi-agents framework simulates a formal debate scenario to engage in discussions on news topics, reaching detection results through debate reasoning and interpretability. 
    }
    \label{fig:intro}
\end{figure}

The rapid propagation of fake news through social networks has become pervasive, significantly impacting individuals, vulnerable communities, and society. This widespread dissemination poses considerable challenges for the governance of public opinion~\cite{bollen2011twitter}, influencing elections~\cite{grinberg2019fake}, undermining public health policies~\cite{gupta2013faking}, fairness~\cite{zhang2023fairlisa,zhangtowards,zhang2024enhancing} and threatening social stability~\cite{starbird2014rumors}. Consequently, the urgency to develop effective and interpretable methods for detecting fake news has never been greater~\cite{roth2022vast}.

Traditional fake news detection methods predominantly rely on machine learning models, particularly those based on BERT~\cite{devlin2018bert}, which excel in text analysis~\cite{kaliyar2021fakebert, mosallanezhad2022domain, hu2023learn}. While these models have achieved notable success, they often suffer from low interpretability and limited generalization capabilities, hindering their effectiveness across diverse contexts. Recent approaches have attempted to leverage LLMs by developing specific prompts to generate explanations and detection results~\cite{hu2024bad,liu2025bidev}. However, these methods may not fully utilize LLMs' inherent reasoning abilities and complex cognitive processes, frequently resulting in excessively simplified and premature conclusions.

In the context of truth-seeking, the proverb "\textit{truth becomes clearer through debate}" emphasizes the importance of examining and challenging information from multiple perspectives to arrive at a more accurate understanding~\cite{branham2013debate,khandebating}. The exchange of differing viewpoints in a structured debate often leads to refining and clarifying ideas. However, existing fake news detection models fail to incorporate this approach fully, often limiting their analysis to a single, unchallenged perspective. This limitation prevents a comprehensive evaluation of news content and may overlook detailed signals that are critical for accurate detection.

To address these challenges, we introduce a novel multi-agent system with LLMs named TruEDebate (TED). TED is designed to enhance the interpretability and effectiveness of fake news detection by following the structured methodologies of formal debate. Inspired by the principle that truth emerges more clearly through debate, our approach simulates dynamic and structured debates among multiple agents, each examining the news from different angles. This debate structure follows the Lincoln-Douglas Debate format~\cite{djuranovic2003ultimate}, where agents engage in a series of stages to present, challenge, and refine their arguments. This framework mirrors the human cognitive process of evaluating information through discourse, fostering deep intellectual engagement and comprehensive evaluation of news content.
In this framework, multiple agents, each assigned distinct roles, engage in a structured debate process to evaluate the truth of news items. This debate process enhances the interpretation of the detection process and reflects the human cognitive process of debate, where conflicting viewpoints are analyzed and integrated.

The TED framework comprises two innovative components: the DebateFlow Agents and the InsightFlow Agents. The DebateFlow Agents organize multiple LLM-powered agents into two opposing teams—one supporting and the other opposing the truth of the news. These agents engage in sequential stages of opening, cross-examination rebuttal, and closing statements, effectively simulating a rigorous debate environment. This setup encourages the exploration of differing viewpoints, refining, and clarifying ideas, and addresses the limitation of existing models that analyze information from a single perspective.
Concurrently, the InsightFlow Agents encompass two specialized sub-agents: the Synthesis Agent and the Analysis Agent. The Synthesis Agent is responsible for summarizing the debates and providing a comprehensive viewpoint, ensuring a coherent and thorough conclusion. The Analysis Agent, composed of a role-aware encoder and a debate graph, integrates role embeddings and models the interactions between different debate roles and arguments through an interaction attention mechanism. By capturing the detailed relationships and discourse structures inherent in the debates, the Analysis Agent effectively identifies critical indicators essential for fake news detection.

We conducted comprehensive experiments on two datasets, ARG-EN and ARG-CN, to evaluate the effectiveness of TED. The results indicate that our framework outperforms traditional methods across multiple evaluation metrics. Additionally, TED facilitates structured debate procedures and enhances logical reasoning, contributing to accurate conclusions. By enabling systematic debate and in-depth analysis, our approach advances the development of information systems capable of making more precise and transparent decision-making processes, thereby supporting the creation of responsible information environments.

Our contribution can be summarized in the following ways:

$\bullet$ We introduce TED, a novel and interpretable fake news detection framework based on LLMs. TED not only performs detection through structured debates but also provides interpretable justifications to aid in understanding the results.

$\bullet$ Inspired by Lincoln-Douglas debate theory~\cite{djuranovic2003ultimate}, which holds that ``truth becomes clearer through debate," we design two key components within our framework: the Synthesis Agent and the Analysis Agent. The Synthesis Agent summarizes the debates and synthesizes comprehensive viewpoints, ensuring coherent conclusions. The Analysis Agent, incorporating a role-aware encoder and a debate graph, integrates role embeddings and models the interactions between different debate roles and arguments through an attention mechanism. This structure enables the Analysis Agent to identify critical indicators essential for accurate fake news detection.

$\bullet$ Extensive experiments on the ARG-EN and ARG-CN datasets demonstrate TED's superior performance to traditional methods, showing its ability to make accurate and transparent decisions.
Furthermore, TED also maintains robust performance across different LLM backbones, showcasing its adaptability and generalizability in fake news detection.

\section{Related Work}

\subsection{Fake News Detection}
Fake news detection is an essential step in the fight against misinformation~\cite{guo2021does}. 
Earlier works in this area include the study by~\citet{qian2018neural}, which focuses on the early detection of fake news, considering only the text of news articles available at the time of detection.  
~\citet{yu2017convolutional} applied a convolutional approach for misinformation identification.
As the field evolved, various methods have been introduced to improve detection efficacy. 
For example, ~\citet{jin2022towards} move towards fine-grained reasoning for fake news detection by better reflecting the logical processes of human thinking and enabling the modeling of subtle clues. 
Furthermore, ~\citet{zheng2022mfan} investigated the integration of diverse multi-modal data considering the complexity of their interrelations. 
Recently, with the rise of LLMs, ~\citet{hu2024bad} combined large and small LMs and achieved good improvements using only textual content for fake news detection. The agent workflow based on a large language model designed by ~\citet{li2024large} uses tool learning and knowledge retrieval to detect fake news. ~\citet{liu2024skepticism,liu2024tiny} employs a multi-agent system to simulate the propagation and evolution of fake news.

However, they did not fully utilize the potential of LLMs. Our research enhances the capability of fake news detection by further leveraging a debate system composed of multiple agents with LLMs. In our approach, we do not call any external tools, nor do we require knowledge-based retrieval, making it easily transferable to any fake news detection scenario.

\subsection{Multi-agent Discussion with LLMs}

Despite significant advancements in LLMs, research on multi-agent discussions, where multiple LLMs interactively collaborate to solve problems, is still in its early stages, aiming to leverage collective capabilities beyond individual models.
A key contribution in this field is ChatEval~\cite{chan2023chateval}, which designs a team of multi-agent referees to independently deliberate and assess the quality of responses generated by various models on open-ended questions and standard natural language generation tasks.
Other studies emphasize diversifying agent roles to enrich the debate process~\cite{park2023generative,li2023camel}. ~\citet{liang2023encouraging} further explores how role assignment impacts the effectiveness of multi-agent interactions. ~\citet{chen2023reconcile} introduce the ReConcile framework, which uses different LLMs and a weighted voting system to reach consensus, capturing a broader range of perspectives. ~\citet{du2023improving} employs multi-agent debate to improve the mathematical abilities of LLMs.
Additionally, ~\citet{zhang2024exploring} analyzes multi-agent discussions from a social psychology perspective to understand how social factors influence collaboration and reasoning processes among agents.

However, these works do not truly depict a debate process within their frameworks and cannot analyze problems from multiple perspectives. Building upon these foundations, our work introduces a structured debate framework that leverages multi-agent interactions to enhance the interpretability and effectiveness of fake news detection.

\subsection{LLM-Based Multi-Agent System}
With LLMs' powerful text generation capabilities, integrating LLMs into simulating social dynamics represents a burgeoning field of research, yielding promising results~\cite{kaiya2023lyfe,li2023quantifying,song2024mmac,wu2024foundations}. 
These LLM-based generative agents excel in digital environments, demonstrating proficiency in natural language tasks~\cite{chen2023improving,chen2023topic,yuhan2023unleashing} and reasoning task~\cite{zhang2025visc}.
~\citet{tornberg2023simulating} used LLMs and agent-based modeling to simulate social media, focusing on news feed algorithms and offering real-world insights.
Further, ~\citet{park2022social} demonstrated that LLM-based agents could generate social media content indistinguishable from that produced by humans. Moreover, ~\citet{qian2023communicative} designed ChatDev, a virtual development company that simulates a human development team.

These frameworks that utilize LLM multi-agent systems to solve complex tasks all share a crucial component: the interaction among multiple agents~\cite{hong2023metagpt,talebirad2023multi,chan2023chateval,wu2023autogen,zhang2024large,zhang2024thinking,liu2025mobile,zhang2024sagraph}. However, these works consider only the message passing between different agents. While some claim to engage in multi-agent debates, they do not adequately depict the debate process during interactions. Our work establishes a standard debate scenario, where various roles debate fake news from different perspectives to determine the truth of the news topic. To the best of our knowledge, this is the first instance of a social debate simulation in the context of fake news.

\section{Methodology}

\begin{figure*}[tb]
    \centering
    \includegraphics[width=\linewidth]{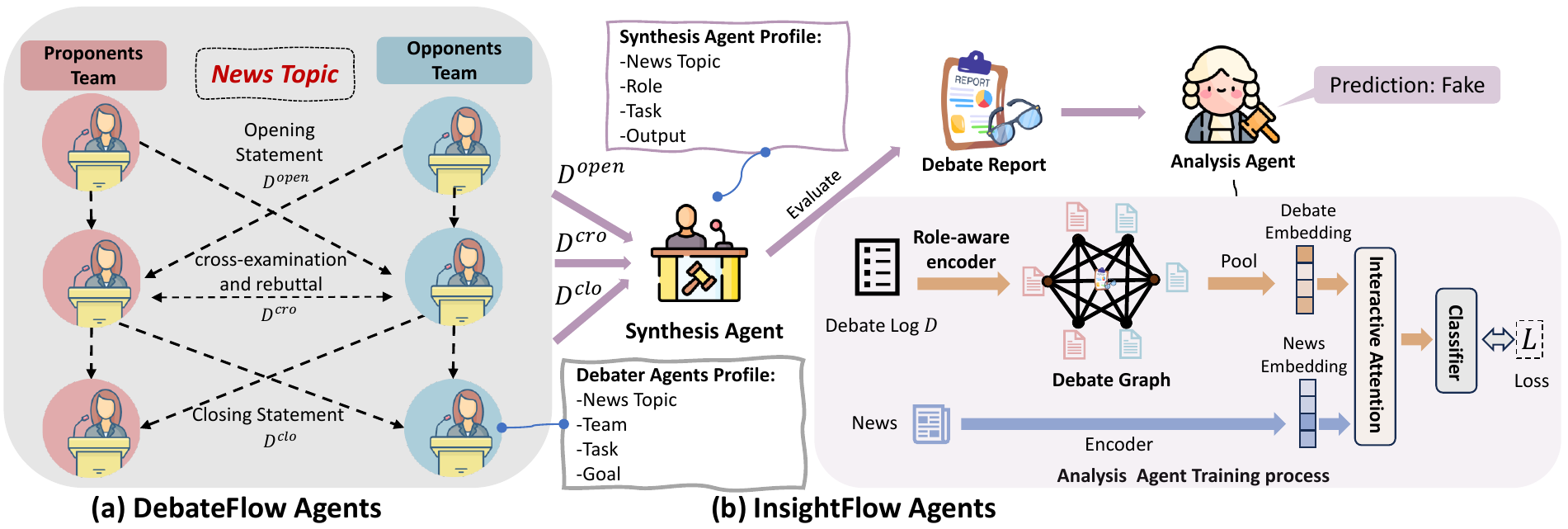}
    \caption{Our TED framework consists of two main components: (a) the DebateFlow Agents and (b) the  InsightFlow Agents. The DebateFlow Agents simulate the reasoning process in a debate scenario, including the Opening, Cross-examination, Rebuttal, and Closing Statement. The InsightFlow Agents consist of the Synthesis Agent and the Analysis Agent. The former generates a debate report based on all debate records, which is then passed to the latter to predict whether the new content is true.}
    \label{fig:model}
\end{figure*}

\subsection{Task Formulation}
We reformulate fake news detection as an interactive debate among multiple LLM agents.
We introduce a system comprising $N$ LLM agents denoted as $\mathcal{A} = ( a_1, \ldots, a_N)$, each assigned unique roles and stances in the debate process.
Given a news item $F$, all agents engage in a structured debate to determine its veracity. The agents are organized into two opposing teams:
\begin{itemize} 
    \item \textbf{Proponents}: Agents who argue that $F$ is true. 
    \item \textbf{Opponents}: Agents who argue that $F$ is fake. 
\end{itemize}
Each agent plays a specific role in the debate and participates in predefined stages. All interactions are recorded in a debate log $D$.
At the conclusion stage of the debate, two types of agents evaluate the outcome. The Synthesis Agent analyzes the debate log $D$ to produce a comprehensive summary report $S$. At the same time, the Analysis Agent utilizes the summary report $S$ and the original news item $F$ to predict the truth value.

Our objective is to leverage LLMs' reasoning capabilities within this debate framework to enhance the effectiveness and interpretability of fake news detection.

\subsection{Our Multi-Agents Debate Framework}
We introduce \textbf{TruEDebate (TED)}, a framework that enhances fake news detection by simulating a structured debate among multiple LLM-powered agents. TED consists of two main components:

\noindent $\bullet$ \textbf{DebateFlow Agents}: Organized into two opposing teams—the \textit{Proponents}, who argue that the news item is true, and the \textit{Opponents}, who argue that it is fake. Inspired by Lincoln-Douglas debate theory~\cite{djuranovic2003ultimate}, these agents engage in a multi-stage debate process, including opening, cross-examination, rebuttal, and closing statements. This setup mirrors formal debate settings, encouraging thorough examination of the news from multiple perspectives.

\noindent $\bullet$ \textbf{InsightFlow Agents}: Comprising the \textit{Synthesis Agent} and the \textit{Analysis Agent}. The Synthesis Agent summarizes the debate, providing a cohesive overview of the key arguments and counterarguments. The Analysis Agent employs a role-aware encoder and a debate graph to model the interactions and discourse structures within the debate, ultimately predicting the authenticity of the news item.

 As shown in Figure~\ref{fig:model}, through the interactions of these agents, TED fosters deep intellectual engagement and logical reasoning, leading to more accurate and interpretable fake news detection. The structured debate process not only refines the agents' understanding of the news item but also provides transparent justifications for the final decision, embodying the principle that ``truth becomes clearer through debate."
\subsection{DebateFlow Agents}

The \textbf{DebateFlow Agents} form the core of the TED framework, orchestrating a structured debate to assess the authenticity of a news item $F$. The agents are divided into two opposing teams: the \emph{Proponents} ($\mathcal{T}_{pro}$), who argue that $F$ is true, and the \emph{Opponents} ($\mathcal{T}_{opp}$), who contend that $F$ is fake. Each team consists of agents assigned specific roles corresponding to different debate stages.

Inspired by Lincoln-Douglas debate~\cite{djuranovic2003ultimate}, which is commonly used in debate settings to resolve controversial issues, the debate is structured into three main stages: \emph{Opening}, \emph{Cross-examination and Rebuttal}, and \emph{Closing}. Let $\Phi=\{\phi_1, \phi_2, \phi_3\}$ denote these stages. In each stage $\phi_k \in \Phi$, agents from both teams participate according to their roles.
In the \emph{Opening} stage ($\phi_1$), agents present their initial arguments supporting their team's stance. Each agent $a_i$ generates an argument $d_i^{(1)}$ based on the news item $F$ and their team's position:

\begin{equation}
    d_i^{(1)} = f_{arg}(F, Stance_i),
\end{equation}
where $f_{arg}$ represents the argumentation prompt function implemented by the LLM, and $Stance_i \in \{True, Fake\}$ corresponds to the team's stance.

In the \emph{Cross-examination and Rebuttal} stage ($\phi_2$), agents challenge the opposing team's arguments through questions and rebuttals. An agent $a_j$ from one team formulates a question or rebuttal $d_j^{(2)}$ targeting the arguments made in the previous stage by the opposing team:
\begin{equation}
    d_j^{(2)} = f_{reb}(D^{(1)}, Stance_j),
\end{equation}
where $f_{reb}$ is the Cross-examination and Rebuttal prompt function, and $D^{(1)}$ is the set of arguments in stage $\phi_1$.

In the \emph{Closing} stage ($\phi_3$), agents summarize their team's position and reinforce key points. Each agent $a_k$ produces a concluding statement $d_k^{(3)}$:
\begin{equation}
    d_k^{(3)} = f_{clo}(D^{(1)} \cup D^{(2)}, Stance_k),
\end{equation}
where $f_{clo}$ is the closing prompt function, and $D^{(1)} \cup D^{(2)}$ represents the accumulated debate content from previous stages.

The debate log $D$ accumulates all interactions during the debate:
\begin{equation}
    D = \bigcup_{k=1}^{3} D^{(k)}, \quad \text{with} \quad D^{(k)} = \{d^{(k)}_i \mid a_i \in \mathcal{A}^{(k)}\},
\end{equation}
where $\mathcal{A}^{(k)}$ is the set of agents participating in stage $\phi_k$.

Each agent's contribution $d^{(k)}_i$ depends on the news item $F$, their team's stance, their role in the debate, and the prior debate context $D^{<k}$:
\begin{equation}
    d^{(k)}_i = f_{\text{LLM}}(F, D^{<k}, \text{Role}_i, \text{Stance}_i),
\end{equation}
where $D^{<k} = \bigcup_{m=1}^{k-1} D^{(m)}$ represents the debate history up to stage $\phi_{k-1}$, and $\text{Role}_i$ specifies the agent's role.

The entire algorithm process can be described in Algorithm~\ref{alg:TruEDebate}. This structured interaction ensures that agents consider both the news content and the evolving debate context, promoting coherence and depth in their reasoning. By engaging in this process, the DebateFlow Agents simulate a rigorous examination of the news item $F$, mirroring the dynamics of human debates and fostering a comprehensive evaluation from multiple perspectives.

The whole set of function prompts can be found in our code files.

\begin{algorithm}[tb]
\caption{TruEDebate (TED): Multi-Agent Debate Framework for Fake News Detection}
\label{alg:TruEDebate}
\begin{algorithmic}[1]
\STATE \textbf{Input:} News item $F$
\STATE \textbf{Output:} Predicted truth value $\hat{y}$ of news item $F$, Explanatory reasons $R$
\STATE \textbf{Initialize DebateFlow Agents:}
\STATE Assign agents into two teams:
\begin{itemize}
    \item \textit{Proponents}: Support that $F$ is true
    \item \textit{Opponents}: Argue that $F$ is fake
\end{itemize}
\STATE Assign roles within each team (e.g., Opening Speaker, Questioner, Closing Speaker)
\STATE \textbf{DebateFlow Agents:}
\FOR{each debate stage $\phi$ in [Opening Statement, Cross-examination, Rebuttal, 
 and Closing Statement]}
        \STATE Agent $i$ presents arguments or rebuttals based on the role
        \STATE Record interactions in debate log $D_{\phi}$
\ENDFOR

\STATE \textbf{InsightFlow Agents:}
\STATE \textbf{Synthesis Agent:}
\STATE Analyze debate logs $D = \{D_{\phi}\}$ to produce evaluation report $S$
\STATE \textbf{Analysis Agent:}
\FOR{each debate interaction (node) $n_i$ in $D$}
    \STATE Obtain node $n_i$ embedding $h_i$ with a Role-aware Encoder
\ENDFOR
\STATE Build Dabate graph $G = (V, E)$ with nodes $V = \{h_i\}$
\STATE Apply GATs to $G$ to obtain graph embedding $g$
\STATE Encode news item $F$ to obtain content embedding $e_F$
\STATE Combined representation $c \gets \text{InteractiveAttention}(g, e_F)$
\STATE Predict truth value $\hat{y}$ using $c$

\RETURN Predicted truth value $\hat{y}$, explanatory reasons $R = \{S, D\}$
\end{algorithmic}
\end{algorithm}

\subsection{InsightFlow Agents}

After the debate conducted by DebateFlow Agents, the \textbf{InsightFlow Agents} analyze the accumulated discourse to produce a final prediction on the authenticity of the news item $F$. This component consists of two principal agents: the \textbf{Synthesis Agent} and the \textbf{Analysis Agent}.

\subsubsection{\textbf{Synthesis Agent}}
The Synthesis Agent processes the debate log $D$, which contains all interactions from the debate stages, to generate a concise and comprehensive summary $S$ of the key arguments and counterarguments presented by both teams. This summary is evaluated from multiple perspectives of fake news detection to ensure a well-rounded judgment. Formally, the core function prompt of the Synthesis Agent is defined as follows:
\begin{tcolorbox}[colback=gray!20, left=1mm, right=1mm, top=1mm, bottom=1mm] 

It should contain a detailed explanation of your assessment of the debate. Focus on evaluating the authenticity of the news involved in the topic by checking the following:

    1. Whether the news contains specific details and verifiable information.
    
    2. Whether the news cites reliable sources or news organizations.
    
    3. The tone and style of the news, with real news generally being more objective and neutral.
    
    4. Any use of emotional language, which might be a characteristic of fake news.
    
    5. Whether the information in the news can be confirmed through other reliable channels.
\end{tcolorbox}

The purpose of the Synthesis Agent is to encapsulate the critical points of the debate, providing a coherent overview highlighting the strengths and weaknesses of each side's arguments. This summary $S$ bridges the debate discourse and the subsequent analytical processes, ensuring that the most pertinent information is readily accessible for further evaluation.

\subsubsection{\textbf{Analysis Agent}}

The Analysis Agent integrates the debate interactions with the original news content to predict the truth value $\hat{y}$ of the news item $F$. This agent employs a combination of role-aware encoder, debate graph, and a news-debate interactive attention mechanism to capture both semantic content and the relational dynamics of the debate, as well as their relevance to the news content.

\paragraph{\textbf{Role-aware Encoder}}

To capture the distinct characteristics of different debate roles, we design a role-aware encoder. For each debate interaction \( d_i \), the node representation is constructed by concatenating the encoder's \texttt{[CLS]} token embedding with a projected role embedding:
\begin{equation}
\begin{aligned}
\mathbf{h}_i^{\text{enc}} &= \text{Encoder}(\text{Text}_i), \\
\mathbf{r}_i^{\text{proj}} &= \mathbf{W}_{\text{role}} \, \mathbf{e}_i,\\
\mathbf{h}_i^{\text{node}} &= \left[\, \mathbf{h}_i^{\text{enc}} \,;\,\mathbf{r}_i^{\text{proj}} \,\right],
\end{aligned}
\end{equation}

\noindent where \( \mathbf{h}_i^{\text{enc}} \in \mathbb{R}^{d_h} \) is the \texttt{[CLS]} token representation from the encoder, \( \mathbf{e}_i \in \mathbb{R}^{d_r} \) is the embedding for role \( r_i \), \( \mathbf{W}_{\text{role}} \in \mathbb{R}^{d_h \times d_r} \) projects the role embedding to match the encoder's dimension, and $[\,;\,]$  denotes vector concatenation.
The resulting node representation \( \mathbf{h}_i^{\text{node}} \) integrates both semantic content and role-specific information of each debate interaction.

\paragraph{\textbf{Debate Graph}}

The encoded node representations are structured into a debate graph \( G = (V, E) \),
where the nodes \( V = \{h_i^{\text{node}}\} \) represent the debate interactions, and the
edges \( E \) represent the relationships between interactions, such as
sequential order or explicit references.

We employ Graph Attention Networks (GAT) \cite{velivckovic2018graph} to
process the node representations. The GAT layers update each node's
representation by attending to its neighbors:
\begin{equation} 
\mathbf{h}_i^{(l+1)} = \sigma\left( \sum_{j \in \mathcal{N}(i)} \alpha_{ij}^{(l)} \mathbf{W}^{(l)} \mathbf{h}_j^{(l)} \right), 
\end{equation}

\noindent where \( \mathbf{h}_i^{(l)} \) is the representation of node \( i \) at layer \( l \), \( \mathcal{N}(i) \) denotes the set of
neighbors of node \( i \), \( \mathbf{W}^{(l)} \) is a learnable weight matrix, \( \sigma \) is an activation
function, and \( \alpha_{ij}^{(l)} \) are the attention coefficients.
By stacking multiple GAT layers, the model captures higher-order interactions and the relational dynamics within the debate.

\paragraph{\textbf{Debate-News Interactive Attention}}

To integrate the debate representations with the news content, we employ an interactive attention mechanism. First, we encode the news content \( F \) and apply global pooling to obtain the graph-level debate representation \( \mathbf{g} \):
\begin{equation}
\begin{aligned}
&\mathbf{e}_F = \text{Encoder}(F), \\
&\mathbf{g} = \text{GlobalPool}(\mathbf{H}^{(L)}),
\end{aligned}
\end{equation}

\noindent where \( \mathbf{H}^{(L)} \) is the output of the \( L \)-th GAT layer, and \texttt{GlobalPool} denotes a global mean pooling operation.

Next, we project both the debate representation and the news content embedding to a common dimensionality:
\begin{equation}
\begin{aligned}
\mathbf{g}^{\text{proj}} &= \mathbf{W}_g \mathbf{g}, \\ \mathbf{e}_F^{\text{proj}} &= \mathbf{W}_e \mathbf{e}_F,
\end{aligned}
\end{equation}
where \( \mathbf{W}_g \) and \( \mathbf{W}_e \) are projection matrices.

We then compute the interaction representation \( \mathbf{c} \) using multi-head attention (MHA) to capture the interactions between the debate and news content:
\begin{equation}
\mathbf{c} = \text{MHA}(\mathbf{e}_F^{\text{proj}}, \mathbf{g}^{\text{proj}}, \mathbf{g}^{\text{proj}}),
\end{equation}
Finally, we concatenate the projected debate representation with the interaction representation to obtain the comprehensive representation:
\begin{equation}
\mathbf{h} = [\mathbf{g}^{\text{proj}}; \mathbf{c}].
\end{equation}
This interactive attention allows the model to focus on different aspects of the debate relevant to the news content, effectively integrating information from both sources.

\paragraph{\textbf{Agent Training}}

The combined representation \( \mathbf{h} \) is processed through a fully connected layer followed by a softmax activation to generate the predicted probability distribution over classes. The model is trained by minimizing the cross-entropy loss between the predicted probabilities and the actual labels:
\begin{equation}
\begin{aligned}
\hat{\mathbf{y}} &= \text{softmax}(\mathbf{W}_{\text{fc}} \mathbf{h} + \mathbf{b}_{\text{fc}}), \\
\mathcal{L} &= - \sum_{k} y_k \log \hat{y}_k,
\end{aligned}
\end{equation}
where \( \mathbf{W}_{\text{fc}} \) and \( \mathbf{b}_{\text{fc}} \) denote the weights and bias of the classification layer, \( \hat{\mathbf{y}} \) represents the predicted probability distribution, \( y_k \) is the ground truth label for class \( k \), and \( \hat{y}_k \) is the predicted probability for class \( k \).

\section{Experiment}

\begin{table*}[ht]
\centering
\caption{The performance of the TED, as well as the LLM-only, SLM-only, and LLM+SLM methods was assessed. All results (except for those of \texttt{GPT-4o-mini} and ChatEval) for the baselines are taken from~\citet{hu2024bad}. The top two results for both macF1-score and accuracy are highlighted, with the highest in each category being \textbf{bolded} and the second highest {\ul underlined}. Numbers in \textbf{bold} mean that the improvements to the baseline models are statistically significant (t-test with p-value$<$0.01).}
\begin{tabular}{@{}clllllllll@{}}
\toprule
\multicolumn{2}{c}{\multirow{2}{*}{\textbf{Model}}}     & \multicolumn{4}{c}{\textbf{ARG-EN}}                                                                                                & \multicolumn{4}{c}{\textbf{ARG-CN}}                                                                                                \\ \cmidrule(lr){3-6} \cmidrule(l){7-10}
\multicolumn{2}{c}{}                                    & \multicolumn{1}{c}{macF1} & \multicolumn{1}{c}{Acc.} & \multicolumn{1}{c}{$F1_\text{real}$} & \multicolumn{1}{c}{$F1_\text{fake}$} & \multicolumn{1}{c}{macF1} & \multicolumn{1}{c}{Acc.} & \multicolumn{1}{c}{$F1_\text{real}$} & \multicolumn{1}{c}{$F1_\text{fake}$} \\ \midrule
\multirow{2}{*}{LLM-Only}        & GPT-3.5-turbo        & 0.702                     & 0.813                    & 0.884                                & 0.519                                & 0.725                     & 0.734                    & 0.774                                & 0.676                                \\
                                 & GPT-4o-mini          & 0.691                     & 0.845                    & 0.909                                & 0.472                                & 0.725                     & 0.746                    & 0.780                                & 0.670                                \\ \midrule
\multirow{4}{*}{SLM-Only}        & Bert             & 0.765                     & 0.862                    & 0.916                                & 0.615                                & 0.753                     & 0.754                    & 0.769                                & 0.737                                \\
                                 & EANN                 & 0.763                     & 0.864                    & 0.918                                & 0.608                                & 0.754                     & 0.756                    & 0.773                                & 0.736                                \\
                                 & Publisher-Emo        & 0.766                     & 0.868                    & 0.920                                & 0.611                                & 0.761                     & 0.763                    & 0.784                                & 0.738                                \\
                                 & ENDEF                & 0.768                     & 0.865                    & 0.918                                & 0.618                                & 0.765                     & 0.766                    & 0.779                                & 0.751                                \\ \midrule
\multirow{4}{*}{LLM+SLM}         & Bert + Rationale & 0.777                     & 0.870                    & 0.921                                & 0.633                                & 0.767                     & 0.769                    & 0.787                                & 0.748                                \\
                                 & SuperICL             & 0.736                     & 0.864                    & 0.920                                & 0.551                                & 0.757                     & 0.759                    & 0.779                                & 0.734                                \\
                                 & ARG                  & {\ul 0.790}               & {\ul 0.878}              & {\ul 0.926}                          & {\ul 0.653}                          & {\ul 0.784}               & {\ul 0.786}              & {\ul 0.804}                          & {\ul 0.764}                          \\
                                 & ARG-D                & 0.778                     & 0.870                    & 0.921                                & 0.634                                & 0.771                     & 0.772                    & 0.785                                & 0.756                                \\ \midrule
\multirow{3}{*}{Multi-Agents} & ChatEval(One-by-One) & 0.733 & 0.860 & 0.919 & 0.546 & 0.694 & 0.717 & 0.778 & 0.611 \\
                                 & ChatEval(Simultaneous-Talk) & 0.738 & 0.869 & 0.923 & 0.553 & 0.694 & 0.719 & 0.780 & 0.608 \\
                                 & \textbf{TED(Ours)} & \textbf{0.803} & \textbf{0.892} & \textbf{0.932} & \textbf{0.674} & \textbf{0.795} & \textbf{0.798} & \textbf{0.815} & \textbf{0.774} \\ \bottomrule
\end{tabular}
\label{table:main}
\end{table*}

\subsection{Experiment Setup}

\subsubsection{Datasets.}
For evaluation, we utilized two datasets: ARG-EN and ARG-CN~\cite{hu2024bad}. These datasets are derived from the Chinese dataset Weibo21 \cite{nan2021mdfend} and the English dataset GossipCop \cite{shu2020fakenewsnet}.
In previous work~\cite{hu2024bad}, they processed these datasets by deduplication and dividing the data temporally. This approach helps prevent overestimating the SLM performance due to potential data leakage. We used these two datasets in the experiment. The statistical details of the datasets are summarized in Table~\ref{tab:fake_news_stats}.

\begin{table}[ht]
\centering
\caption{Statistics of the fake news detection datasets.}
\begin{tabular}{@{}ccccccc@{}}
\toprule
 & \multicolumn{3}{c}{\textbf{ARG-CN}} & \multicolumn{3}{c}{\textbf{ARG-EN}} \\ 
 \cmidrule(r){2-4} \cmidrule(l){5-7}
 & \textbf{Train} & \textbf{Val} & \textbf{Test} & \textbf{Train} & \textbf{Val} & \textbf{Test} \\ 
\midrule
\textbf{Real} & 2,331 & 1,172 & 1,137 & 2,878 & 1,030 & 1,024 \\ 
\textbf{Fake} & 2,873 & 779 & 814 & 1,006 & 244 & 234 \\ 
\textbf{Total} & 5,204 & 1,951 & 1,951 & 3,884 & 1,274 & 1,258 \\ 
\bottomrule
\end{tabular}
\label{tab:fake_news_stats} 
\end{table}

\begin{table*}[htbp]
\centering
\caption{Ablation study of our model TED on ARG-EN and ARG-CN datasets. Numbers in \textbf{bold} mean that the improvements to the ablation models are statistically significant (t-test with p-value$<$0.01).}
\begin{tabular}{@{}lllllllll@{}}
\toprule
\multicolumn{1}{c}{\multirow{2}{*}{}}               & \multicolumn{4}{c}{\textbf{ARG-EN}}                                                                                         & \multicolumn{4}{c}{\textbf{ARG-CN}}                                                                                         \\ \cmidrule(r){2-5} \cmidrule(l){6-9} 
\multicolumn{1}{c}{}                                & \multicolumn{1}{c}{macF1} & \multicolumn{1}{c}{Acc.} & \multicolumn{1}{c}{$F1_\text{real}$} & \multicolumn{1}{c}{$F1_\text{fake}$} & \multicolumn{1}{c}{macF1} & \multicolumn{1}{c}{Acc.} & \multicolumn{1}{c}{$F1_\text{real}$} & \multicolumn{1}{c}{$F1_\text{fake}$} \\  \midrule
\multicolumn{1}{l}{\textbf{Full Model (TED)}} & \textbf{0.803}            & \textbf{0.892}           & \textbf{0.932}               & \textbf{0.674}               & \textbf{0.795}            & \textbf{0.798}           & \textbf{0.815}               & \textbf{0.774}               \\
 \quad \textit{w/o DebateFlow Agents}                 & 0.780                     & 0.874                    & 0.918                        & 0.642                        & 0.767                     & 0.781                    & 0.795                        & 0.739                        \\
 \quad \textit{w/o Synthesis Agent}                & 0.792                     & 0.881                    & 0.924                        & 0.660                       & 0.771                     & 0.787                    & 0.801                        & 0.741                        \\
 \quad \textit{w/o Analysis  Agent}                & 0.729                     & 0.856                    & 0.915                        & 0.543                        & 0.737                    & 0.736                    & 0.785                        & 0.689                        \\ \bottomrule
\end{tabular}
\label{table:ablation}
\end{table*}

\subsubsection{Baselines.}
Following previous work~\cite{hu2024bad}, we conducted experiments comparing three types of methods.

\noindent $\bullet$ \textbf{LLM-Only Methods}: Directly use GPT-3.5-turbo and GPT-4o-mini as a LLM for fake news detection.

\noindent $\bullet$ \textbf{SLM-Only Methods}: Small Language Models (SLMs), like BERT-based models, generally perform well on fake news detection tasks. In this category, we include the following approaches:

\textbf{(1) Bert}~\cite{devlin2018bert}: Detecting fake news using a fine-tuned vanilla BERT-base model. 

\textbf{(2) EANN}~\cite{wang2018eann}: This model employs auxiliary adversarial training to effectively isolate and reduce event-related features, using the publication year as the label for the auxiliary task.

\textbf{(3) Publisher-Emo}~\cite{zhang2021mining}: A model that integrates a range of emotional attributes with textual features to detect fake news.

\textbf{(4) ENDEFA}~\cite{zhu2022generalizing}: A model that employs causal learning to eliminate entity bias, enhancing its generalization capabilities on fake news datasets with distribution shifts. All approaches in this category utilize the same SLM (BERT) for text encoding.

\noindent $\bullet$ \textbf{LLM+SLM Methods}. LLM solely often does not perform well on the task. Hence, a number of works combine LLM with SLM to enhance performance.

\textbf{(1) Bert+Rationale}: It combines features from both the news and the rationale encoder, feeding them into an MLP for prediction.

\textbf{(2) SuperICL}~\cite{xu2023small}: It utilizes the SLM as a plug-in to enhance the in-context learning capabilities of the LLM by incorporating predictions and confidence levels for each test sample into the prompt.

\textbf{(3) ARG}~\cite{hu2024bad}: This model designs an Adaptive Rationale Guidance network for fake news detection, where SLMs selectively derive insights on news analysis from the rationales provided by LLMs.

\textbf{(4) ARG-D}~\cite{hu2024bad}: It is a rationale-free version of ARG created through distillation, designed for cost-sensitive scenarios that do not require querying LLMs.

\subsubsection{Implementation Details.}
We use Python to operationalize our framework TED. The LLM used is \texttt{GPT-4o-mini} accessed via OpenAI API calls due to its representativeness and convenient calling. 
Agents and the scene they live in were defined using a Python library Mesa~\cite{kazil2020utilizing}.
For Analysis Agent, We utilize a fine-tuned BERT model due to its extensive application in this domain as highlighted by recent studies~\cite{kaliyar2021fakebert}. For our experiments, we employ models from the Transformers package \textit{chinese-bert-wwm-ext} for Chinese and \textit{bert-base-uncased} for English evaluations. Optimization is performed using Adam~\cite{kingma2014adam}, with a comprehensive grid search to determine the optimal learning rate. Performance is evaluated using the best-validation checkpoint from the testing results. For the Analysis Agent, the experiments were conducted using PyTorch, with the HuggingFace Transformers library for pre-trained models and PyTorch Geometric for graph neural network components. 
Text data was tokenized using the BERT tokenizer appropriate for the dataset's language. Debate interactions were encoded with role-aware embeddings, and news content was separately encoded using the BERT model.
Additionally, a fixed graph structure was employed for each debate instance, with nodes representing debate interactions and edges capturing the logical flow and relationships between these interactions. Edges were predefined based on the debate sequence and interactions between roles.
All experiments were run on machines equipped with NVIDIA 3090 GPUs to accelerate training.
The specific experimental implementation details can be found in our code.

\subsection{Main Results}

Table~\ref{table:main} presents the performance of our proposed TED framework and various other models, such as LLM-only, SLM-only, and LLM+SLM methods, on the ARG-EN and ARG-CN datasets. From the results, we observe:
(1) TED outperforms all other methods, achieving the highest macF1 scores and accuracy across the datasets, showcasing its exceptional capability in detecting fake news. This underlines the efficacy of our multi-agent debate method in thoroughly analyzing and judging the truth of news items.
(2) The LLM-only method, which utilizes LLMs without structured argumentation, exhibits inferior performance compared to TED. This emphasizes the limited capability of LLMs when deployed in isolation, lacking the structured interaction and discourse mechanisms provided in TED.
(3) Comparisons among SLM-only methods reveal incremental improvements across each variant, yet none achieve the performance level of TED. This suggests that while traditional SLMs are capable, they fall short of the robust and comprehensive analytical framework offered by the multi-agent system in TED.
(4) The LLM+SLM methods, attempting to combine the capabilities of LLMs with SLMs, show varied performance. For instance, the ARG model demonstrates considerable effectiveness but still does not reach the level of TED. This indicates that while the integration of LLM and SLM provides benefits, the structured debate format of TED is crucial for optimal performance.

In addition, to evaluate the effectiveness of our TED framework for fake news detection, we compared it with the ChatEval framework~\cite{chan2023chateval}, which uses \textit{One-by-One} and \textit{Simultaneous-Talk} strategies for multi-agent debates. Applying these strategies to our fake news detection task, as shown in Table~\ref{table:main}, TED consistently outperforms ChatEval across all metrics. The formal debate structure of TED promotes more interactive reasoning among agents, allowing for a deeper analysis of news content. 
In contrast, ChatEval's strategies do not achieve the same depth in fake news detection. 
These results highlight TED's superiority in enhancing the accuracy and reliability of multi-agent systems for fake news detection.
By fostering structured and dynamic interactions, TED not only improves detection accuracy but also ensures a more thorough examination of the news item.

\section{Analysis and Discussion}

\subsection{Ablation Study}

We conducted an ablation study to assess the impact of each component in the TED framework. The results indicate that removing the DebateFlow Agents reduces macF1 scores by approximately 2.3\% on ARG-EN and 2.8\% on ARG-CN. This decline underscores the importance of the structured debate process in facilitating comprehensive multi-perspective analysis, which is essential for accurately determining the authenticity of news items. Removing the Synthesis Agent results in a moderate decrease in macF1 scores, dropping by 1.1\% on ARG-EN and 2.4\% on ARG-CN. This highlights the critical role of the Synthesis Agent in consolidating debate interactions into coherent summaries, thereby enhancing the model’s ability to integrate diverse arguments effectively for precise classification. Omitting the Analysis Agent causes the most significant drop, with macF1 scores falling by 8.0\% on ARG-EN and 4.5\% on ARG-CN, emphasizing its crucial function in integrating debate insights with the news content. Overall, the ablation results confirm that each component of the TED framework is indispensable for achieving high performance in fake news detection. The DebateFlow Agents enable robust multi-agent interactions; the Synthesis Agent ensures effective consolidation of debate outcomes, and the Analysis Agent integrates these insights with the news item to get precise results.

\subsection{Performance on Different Backbones}

\subsubsection{Performance on closed-source models}
\begin{figure}[ht]
    \centering
    \includegraphics[width=1\linewidth]{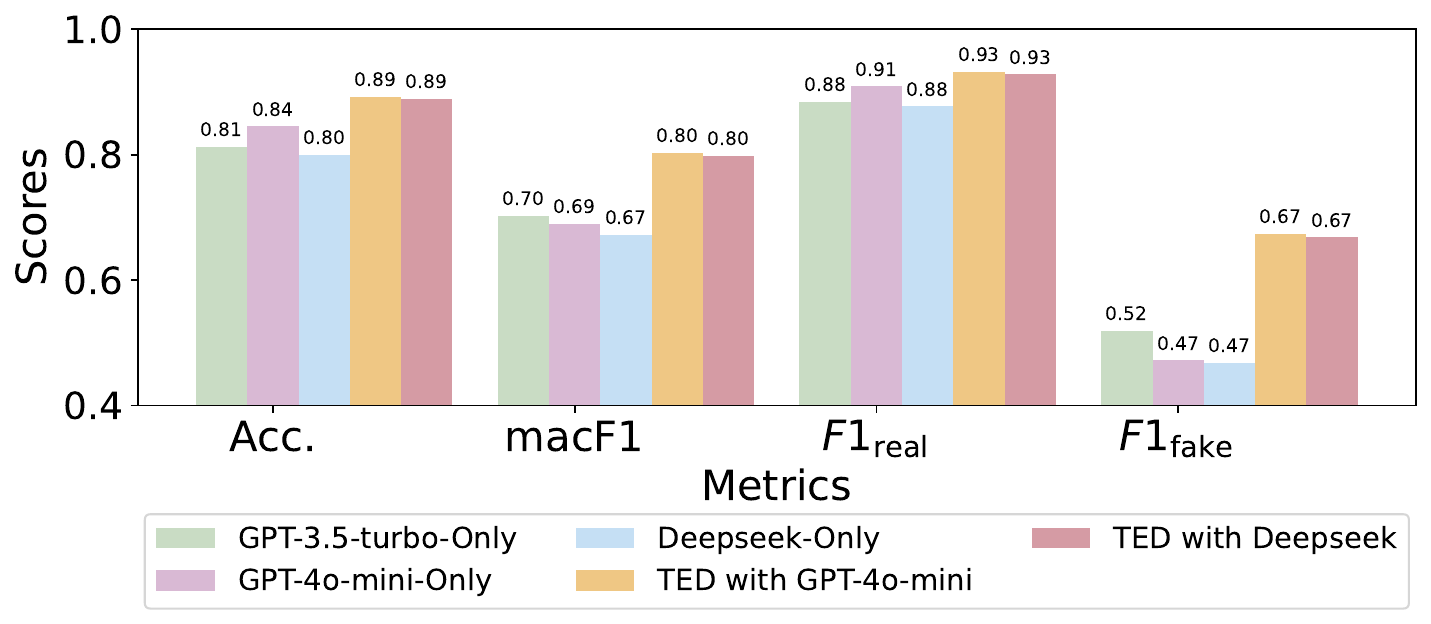}
    \caption{Performance of our framework TED with GPT-4o-mini and Deepseek as closed-source backbones respectively.}
    \label{fig:moon}
\end{figure}

To evaluate the robustness and versatility of the TED framework, we conducted experiments on the ARG-EN dataset by replacing the DebateFlow Agents and Synthesis Agent with different closed-source models, namely GPT-4o-mini and Deepseek\footnote{https://www.deepseek.com/}. Figure~\ref{fig:moon} demonstrates that TED consistently enhances performance compared to the baseline LLM-Only model across all evaluation metrics, regardless of the specific closed-source model used. This outcome highlights the framework's adaptability to different large-scale models, showing that TED effectively leverages the unique strengths of various closed-source backbones.
Additionally, we observed that the performance of TED with GPT-4o-mini and TED with Deepseek was comparable, with a slight advantage in favor of GPT-4o-mini. This suggests that while the two models are competitive, GPT-4o-mini appears to have a slight edge in some aspects, particularly in handling complex reasoning tasks. Despite this, the results across both backbones emphasize TED's robustness in boosting fake news detection performance regardless of the specific closed-source model used.
These findings underline that TED is not limited to a single backbone and can effectively enhance performance across different closed-source models. The slight superiority of TED with GPT-4o-mini reinforces the notion that TED can leverage the inherent strengths of each model's reasoning abilities. This adaptability ensures that TED remains effective and reliable in real-world applications.

Overall, the TED framework proves to be both effective and adaptable, maintaining strong performance across a variety of closed-source models. These results validate TED's utility in diverse application scenarios, making it a highly versatile tool for fake news detection.

\begin{table}[htbp]
\centering
\caption{Performance of TED with Qwen2.5 and Llama3.1 family models as backbones, respectively. Numbers with underlined mean the best performance with an open-source model and are statistically significant(t-test with p-value$<$0.01).}
\begin{tabular}{@{}lcllll@{}}
\toprule
\textbf{Backbones}       & \multicolumn{1}{l}{\textbf{\#Param}} & \textbf{macF1} & \textbf{Acc}   & \textbf{$F1_\text{real}$} & \textbf{$F1_\text{fake}$} \\ \midrule
GPT-4o-mini & -                           & 0.803 & 0.892 & 0.932    & 0.674    \\ \midrule
Qwen 2.5~\cite{yang2024qwen2}    & 7B                          & \textbf{0.794} & \textbf{0.887} & \textbf{0.926}    & \textbf{0.661}    \\
Llama 3.1~\cite{dubey2024llama}   & 8B                          & 0.787 & 0.879 & 0.920    & 0.653    \\
Llama 3.1~\cite{dubey2024llama}   & 70B                         & 0.790 & 0.884 & 0.923    & 0.657    \\ \bottomrule
\end{tabular}
\label{lab:open}
\end{table}

\subsubsection{Performance on open-source models}

To further evaluate the adaptability and robustness of the TED framework, we tested its performance on the ARG-EN dataset using several open-source language models, including Qwen-2.5~\cite{yang2024qwen2} and the Llama 3.1 family~\cite{dubey2024llama}. As demonstrated in Table~\ref{lab:open}, TED consistently improves performance across all the open-source models, highlighting its ability to significantly enhance fake news detection regardless of the specific language model used. This shows that TED is not constrained by the availability or type of backbone model, making it a versatile and effective tool for various real-world applications.
Additionally, TED's performance with Qwen 2.5 slightly outperforms the Llama 3.1 variants. Although the differences in performance are minimal, this suggests that Qwen 2.5 is particularly well-suited to TED's debate-driven framework. It may offer a better handling of complex argumentation and context, which is essential for the structured evaluation TED relies on. On the other hand, TED with the Llama 3.1 models also performs admirably, and all configurations lead to better results than the baseline methods, further validating TED's effectiveness across different models.
Although the performance with open-source models is lower than the results obtained with GPT-4o-mini, this can likely be attributed to differences in the underlying reasoning capabilities of the models. Open-source models, while highly capable, may not exhibit the same level of nuanced reasoning as closed-source models like GPT-4o-mini. Nevertheless, TED continues to produce strong results even with these open-source backbones, demonstrating that it can still leverage each model's strengths to achieve high levels of accuracy and reliable detection.

Overall, the results highlight TED's robustness and versatility. It consistently enhances fake news detection performance. These findings suggest that TED is well-suited for deployment in various settings, including scenarios where access to specific models may be limited or open-source models are preferred.

\subsection{Case Study}

\begin{figure}[ht]
    \centering
    \includegraphics[width=1\linewidth]{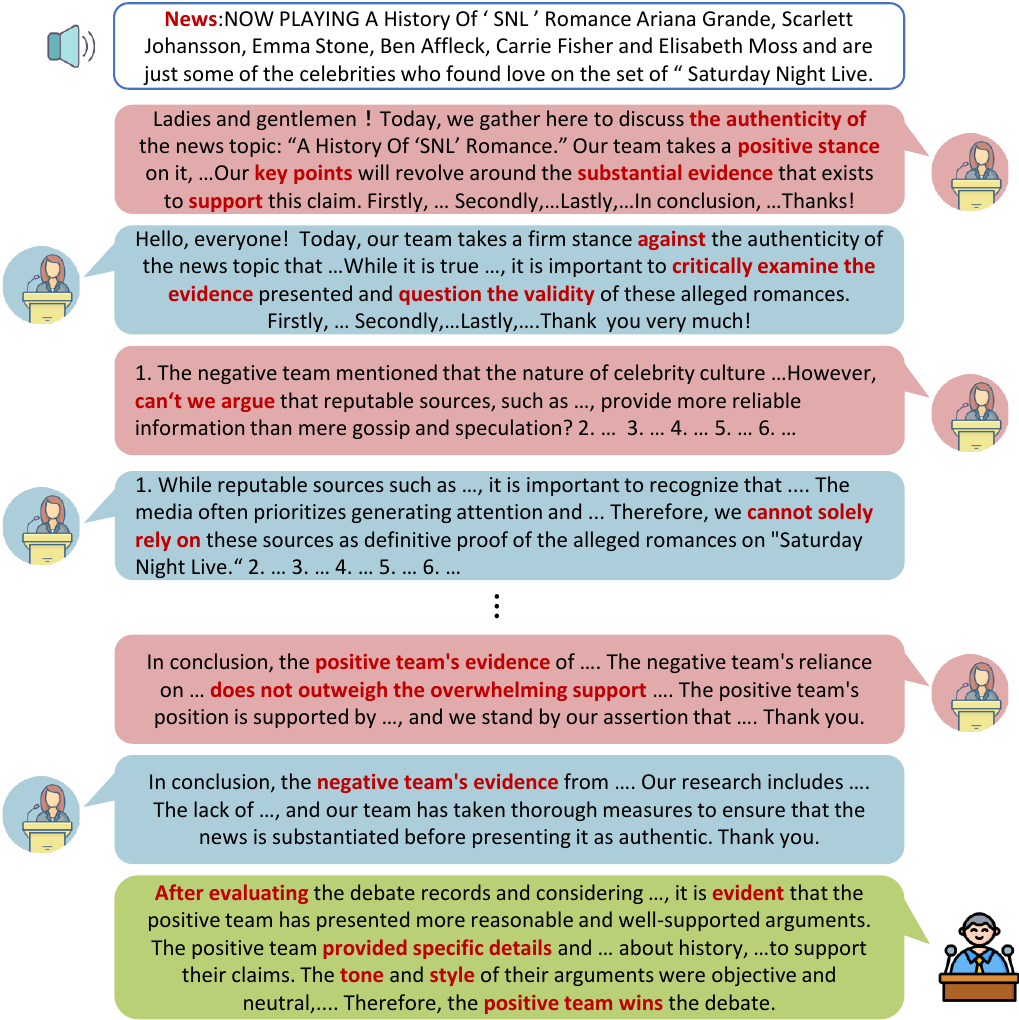}
    \caption{A real debate case study, with its debate report, can serve as justification for interpretable fake news detection.}
    \label{fig:case}
\end{figure}

Figure~\ref{fig:case} presents a comparative micro-analysis of a debate process on fake news as a case study. Participants are divided into pro and con teams in this debate, completing the debate scenario through opening, cross-examinations, and closing statements.
At the beginning of the debate, the pro team clearly stated, ``The primary objective of this debate is to discuss the truth of the news," and they presented key points as evidence to support their stance. The con team adopted a similar approach, presenting their assertions. During the Questioning and Rebuttal phase, both sides challenged each other's viewpoints and responded to the queries raised. In the Closing Statement phase, the pro team reviewed and summarized the prior debate questioned the con team's arguments and reinforced their own points. Although the con team reiterated their position, they did not profoundly challenge the pro team's viewpoints.
Ultimately, based on factors such as the debate's detail, tone, and style, the judges decided to favor the pro team. This entire debate process forms a ``DebateFlow," clearly demonstrating the reasoning process. The judges' debate report also serves as a basis for determining whether news is fake, thus achieving an interpretable fake news detection. Such interpretability is crucial for real-life information verification, representing a significant advantage of the TED framework, an aspect that other models struggle to achieve.

\section{Conclusion}

In conclusion, our study introduces and evaluates the TED framework, a novel approach to fake news detection that incorporates the strengths of LLMs within a structured multi-agent debate. Through extensive experimental validation on the ARG-EN and ARG-CN datasets, TED consistently outperformed traditional machine learning and standalone LLMs across multiple performance metrics. Inspired by human debate, the framework's design leverages diverse perspectives and structured argumentation, providing a robust mechanism for assessing the truth of news items. 
Our work not only advances the field of fake news detection by enhancing accuracy and interpretability but also offers valuable insights for developing responsible information systems. By enabling more reliable and transparent content evaluation, TED contributes to creating information retrieval systems that can better support users in navigating and assessing the growing complexity of digital information.

\appendix

\section{Appendix}
\label{appendix}

In the experiment using the ARG-CN dataset, the prompt we used was the corresponding Chinese version of the aforementioned prompt set. 
And all other function prompts can be found in our code files at https://github.com/LiuYuHan31/TED\_fake-news-detction.

\begin{acks}
This work was supported by the Beijing Outstanding Young Scientist Program (NO. BJJWZYJH012019100020098) and the Intelligent Social Governance Platform, a Major Innovation \& Planning Interdisciplinary Platform for the ``Double-First Class'' Initiative at Renmin University of China. Yuhan Liu is supported by the "Qiushi Academic-Dongliang" Project of Renmin University of
China (No. RUC24QSDL015).
\end{acks}

\bibliographystyle{ACM-Reference-Format}
\balance
\bibliography{sample-base}


\end{document}